\documentclass{aa}
\usepackage{graphicx}
\begin{document}
   \title{On the metal abundance in the core of M87}


   \author{Silvano Molendi
          \inst{1}
          \and
           Fabio Gastaldello
           \inst{1,2}
          }

   \offprints{S.Molendi}

   \institute{
         Istituto di Fisica Cosmica, CNR, via Bassini 15,
         I-20133 Milano, Italy\\
   \and
         Universit\`a di Milano Bicocca, Dip. di Fisica, P.za della Scienza 3
         I-20133 Milano\\
              }


   \abstract{
         We revisit the XMM-Newton  observation of M87
         to make a new and more detailed measurement of the 
         metal abundance profile. 
         After having verified that spectral fits
         with single temperature models show a dramatic abundance
         drop within 1 arcmin of the cluster core,  
         we show that more appropriate models, which include 
         a multi-temperature component to account for the strong
         temperature gradient observed in M87, and a power-law
         component to account for the emission of the nucleus 
         and knot A,  give a substantially flat abundance profile. 
         The  drastic abundance decrement found by fitting a single 
         temperature component to the data is an artifact known 
         as ``$Fe$-bias'' (Buote 2000a,b) following from the application 
         of an oversimplified spectral model to the data. 
   }

   \maketitle
%

\section{Introduction}

Our proximity to the Virgo cluster allows us to study 
its core on physical scales unachievable in other similar systems. 
M87, the giant elliptical galaxy at the 
center of Virgo, is not surprisingly amongst  
the most studied extra-galactic objects in the sky.
At X-ray wavelengths it has been observed by all
major observatories. 
Deprojection of the X-ray images obtained with the
IPC instrument on Einstein (Fabian et al. 1984) showed that the gas 
in M87 has cooling times shorter than the Hubble
time. Further evidence of cooling gas has come from 
the spectrometers on-board Einstein, 
which showed that the gas in M87 was not single-temperature
(e.g. Canizares et al. 1982). Subsequent observations
with ROSAT (Nulsen \& B\"ohringer 1995), ASCA (Matsumoto
et al. 1996) and BeppoSAX (Guainazzi \& Molendi 2000), showed
evidence of a temperature decrement with decreasing radius.
The ASCA and BeppoSAX data also showed evidence of a metal  
abundance increase towards the center of M87 on scales
of a few arcmin, which correspond to a few tens of
kpc. 

A first set of results from the XMM-Newton observation 
of M87 have been recently published in B\"ohringer et al. (2001) 
and Belsole et al. (2001). One of the most striking
results presented is a  
very marked drop in the abundance profiles of $Si$ and $Fe$
within 1 arcmin of the center of M87. B\"ohringer et al. (2001) 
suggest that resonant scattering may be responsible for
this drop. Mathews et al. (2001),   
using detailed radiation transfer  calculations, remark that
the resonant scattering effect is insufficient to explain 
the observed profile and that some sort of continuous opacity 
is required. 

In a recent paper (Molendi \& Pizzolato 2001) we have shown 
that the multi-phase appearance of the EPIC spectra of M87
within 2 arcmin from the core is not ascribable to 
the presence of a truly multi-phase gas but is rather
the consequence of having, within the same line of sight, 
emission from gas at different radii with different temperatures.
 In this Letter we revisit the XMM-Newton  observation of 
M87; our goal is to remeasure the metal abundance profile 
by performing a more advanced spectral analysis  of the EPIC
data with models which take into account the projection effects
leading to the multi-phase appearance of the gas in M87.
 
The remainder of this Letter is organized as follows. 
In section 2 we describe the data preparation. In section 
3 we discuss the spectral models. In section 4 
we present and discuss the results of our analysis.


\section{Observations and Data Preparation}

We use XMM-Newton EPIC data from the PV observation 
of M87/Virgo.
Details on the observation, as well as results from a first 
analysis of this object, have already been 
published in B\"ohringer et al. (2001) 
and Belsole et al. (2001).  

We have obtained calibrated event files for the MOS1, MOS2
and PN cameras with SASv5.0. 
Data were manually screened to remove any remaining bright 
pixels or hot column; for the PN camera we also exclude 
regions contaminated by out of time events. 
Periods in which the background is 
increased by soft-proton flares have been excluded using an 
intensity filter; we rejected all events accumulated  when 
the count rate exceeds 15 cts/100s in the $[10-12]$ 
keV band for the MOS cameras and 25 for the PN.
We have accumulated spectra in 4 concentric annular regions
centered on the emission peak. We extend our analysis 
out to 3 arcmin from the emission peak.
The bounding radii for the regions are: 
 0$^{\prime\prime}$ and 30$^{\prime\prime}$ for region 1;
30$^{\prime\prime}$ and 1$^{\prime}$        for region 2;
1$^{\prime}$        and 2$^{\prime}$        for region 3 and
2$^{\prime}$        and 3$^{\prime}$        for region 4.
We have removed point sources, and the substructures which
are clearly visible from the X-ray image (e.g. Belsole et al.
2001) except in the innermost region. Here we have kept 
the nucleus and knot A, the reason being that on angular scales
this small it is not possible to exclude completely  
their emission. As discussed in section 
3 we prefer to fit the spectrum of this region 
with a model which includes a power-law component
to fit the two point-like sources; note that we include only one power-law
component because the two sources have similar spectra
(B\"ohringer et al. 2001).

Spectra have been accumulated for MOS1 and MOS2 independently.
The Lockman hole observations have been used for the background. 
Background spectra have been accumulated 
from the same detector regions as the source spectra.
The line of sight galactic absorbing column depth towards
the  background field is somewhat smaller,  
$N_{\rm H} \sim 6\times 10^{19}$ cm$^{-2}$, than that found on the line of sight 
of M87. This has a negligible impact on the soft part of our M87 spectra
because the $N_{\rm H} $ variation will lead to a maximum difference of 
10\% in the background intensity, which is in turn never more than 2\% 
of the source intensity in the 0.5-1.0 keV energy range.

The vignetting correction has been applied to the spectra 
rather than to the effective areas, similarly
to what has been done by other authors who 
have analyzed EPIC data (Arnaud et al. 2001).
Spectral fits were performed in the 0.5-4.0 keV band.
Data below 0.5 keV were excluded to avoid residual calibration 
problems in the MOS response matrices at soft energies.
Data above 4 keV were excluded because above this energy the 
spectra show a substantial contamination from hotter gas emitted 
further out in the cluster, on the same line of sight. 

MOS and PN spectra are analyzed separately. As discussed in 
Molendi (2001), there are residual cross-calibration uncertainties
between the two instruments which, as we shall see in the next
sections, sometimes lead to different measurements
of spectral parameters.  


\section{Spectral Modeling}

All spectral fitting has been performed using version 
11.0.1 of the XSPEC package. All models discussed below
include a multiplicative component to account for the
galactic absorption on the line of sight of M87.
The column density is always fixed at a value of
1.8$\times10^{20}$ cm$^{-2}$, which is derived from 21cm measurements
(Lieu et al. 1996).
We have compared our data with three different 
spectral models.

Model I is a single temperature model (vmekal model in XSPEC), 
which allows to fit separately individual metals abundances. 
This model has 13 free parameters: the temperature $T$, the
normalization and the abundance of  $O$, $Ne$, $Na$, $Mg$,
$Al$, $Si$, $S$, $Ar$, $Ca$, $Fe$ and $Ni$, which are all expressed 
in solar units. 

Model II features two single temperature components
(vmekal + vmekal in XSPEC) 
This model has two free parameters more than model I:
the temperature and normalization of the second
thermal component. The metal abundance of each element of 
the second thermal component is linked 
to the same parameter of the first thermal component.
This model has 
been used in the past (e.g. 
Makishima et al. 2001 and refs. therein) as an alternative
to cooling-flow models. 

Model III includes a single temperature  
plus a multi-phase component
 (vmekal + vmcflow in XSPEC). 
This model has two free parameters more than model I:
the minimum temperature, $T_{\rm min}$, and the normalization 
of the multi-phase component.
The other parameters of the multi-phase
component are not free: the maximum temperature, $T_{\rm max}$, and the 
metal abundance of each element are linked respectively to the 
temperature, $T$, and the metal abundances of the same
elements of the single-phase component.
In a recent paper (Molendi \& Pizzolato 2001) we have shown
that such a model fits adequately the EPIC spectra of three clusters
which feature a temperature decrement in their core.
The vmcflow component we use has originally been written
to fit spectra from multi-phase gas, i.e. gas that at
a given physical radius is characterized by a distribution of
temperatures rather than by a single temperature.
Here we are using it to describe a different physical
scenario where, at 
a given physical radius the gas is all at one temperature,
and the multi-phase appearance of the spectrum comes
from having emission from many different physical radii
all within the same line of sight. 

For the spectrum accumulated in the innermost region
models II and III also include a power-law component 
to model the emission of the nucleus and of knot A. 

   \begin{figure}
   \centering
   \includegraphics[angle=-90,width=8.0cm]{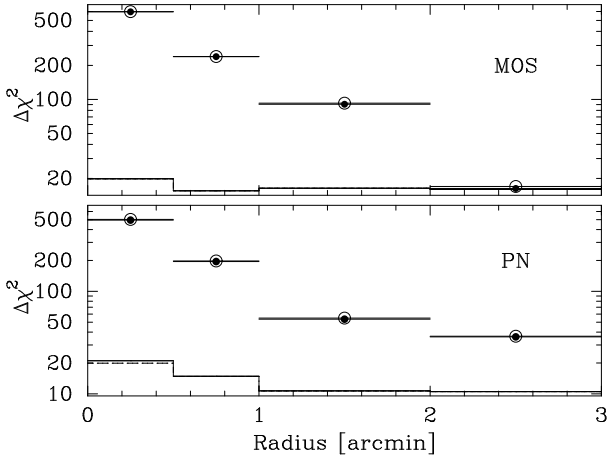}
      \caption{
$\Delta \chi^2$ between models I and II (filled circles), 
and I and III (open circles), as a function of radius.  
The solid (dashed) horizontal line indicates 
the $\Delta \chi^2$ value for which the statistical improvement of 
the model II (III) fit with respect to model I is significant at the 
99\% level according to the F-test.
Top panel is for MOS and bottom for PN.
              }
   \end{figure}
%

   \begin{figure}
   \centering
   \includegraphics[angle=-90,width=8.0cm]{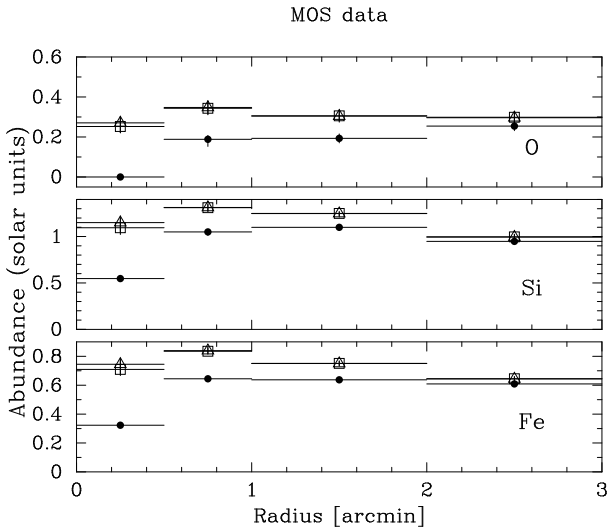}
      \caption{
MOS abundance profiles for O, Si and Fe. 
Uncertainties are at the
68\% level for one interesting parameter ($\Delta \chi^2 = 1$). 
Full symbols indicate the  measurements with model I, empty squares  
and empty triangles those for  models II and III respectively.
              }
   \end{figure}
%

   \begin{figure}
   \centering
   \includegraphics[angle=-90,width=8.0cm]{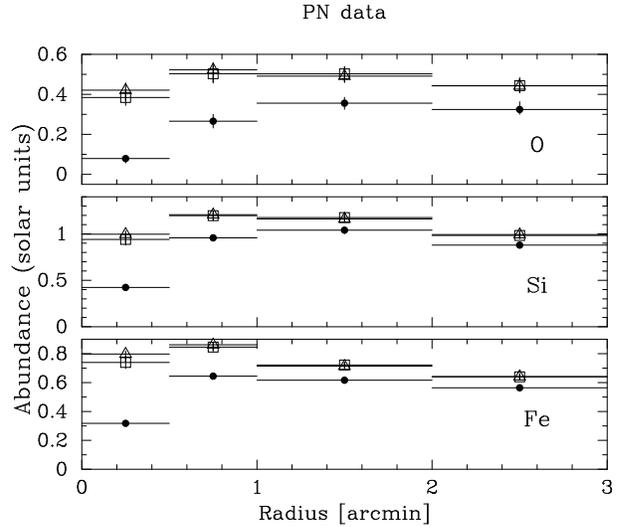}
      \caption{
PN profiles for O, Si and Fe. 
Symbols as in Fig. 2.
              }
   \end{figure}
%

\section{Results}

In  Figure 1 we plot the  
$\Delta \chi^2$ between models I and II (filled circles) 
and models I and III (open circle) as a function of radius.  
The solid (dashed) horizontal line indicates 
the $\Delta \chi^2$ value for which the statistical improvement of 
the model II (III) fit with respect to model I is significant at the 
99\% level according to the F-test.
The results plotted in the top panel refer to the MOS 
spectra while those shown in the bottom panel are for 
the PN spectra.
From Figure 1 we see that, for all bins and for both MOS and PN,
the improvement  of models II and III with respect to model I is significant 
at more than the 99\% level, implying that the 
spectra are characterized by more than just one temperature.

In the case of the innermost bin we have verified that including
only the second thermal component for models II and III,  provides
a significantly better fit with respect to model I and that the
further inclusion of the power-law component is also statistically 
significant. Thus for this bin we have evidence in favor of a second thermal
component and of a power-law component.


In Figures 2 and 3 we report respectively the MOS and PN radial abundance 
profiles of $O$ (top panel), $Si$ (middle panel) and $Fe$ (bottom panel),
while in  Figures 4 and 5 we show those for  
$Mg$ (top panel), $S$ (middle panel) and $Ar$ (bottom panel). 
Full symbols indicate the 
measurements with model I, empty squares and triangles those 
with models II and III respectively.

Figures 2, 3, 4 and 5 show that, for the two innermost
bins, all the metal abundances measured with model I, either with MOS or 
PN, are systematically smaller than those measured with models II
or III. 
The difference between the measurements 
becomes larger for smaller radii as the significance of the improvement
of models  II and III with respect to model I (see Figure 1) increases.

We note that, while for $Si$, $Fe$, $S$ and $Ar$ the measurements obtained 
with PN and MOS are consistent, for $O$ and $Mg$ abundances derived 
with the MOS are systematically higher than those derived with the PN.
The reason lies in the current cross-calibration uncertainties 
between the spectral response of the two instruments (Molendi 2001).
However, the shape of the abundance profiles 
derived with models II and III is the same for MOS and PN: neither 
shows evidence of a dramatic metal abundance drop such as is observed
with model I.

   \begin{figure}
   \centering
   \includegraphics[angle=-90,width=8.0cm]{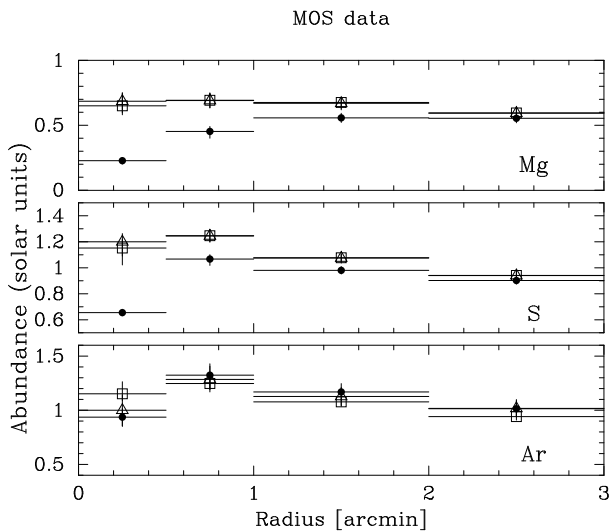}
      \caption{
MOS profiles for Mg, S and Ar. Symbols as in Fig. 2. 
              }
   \end{figure}
%

   \begin{figure}
   \centering
   \includegraphics[angle=-90,width=8.0cm]{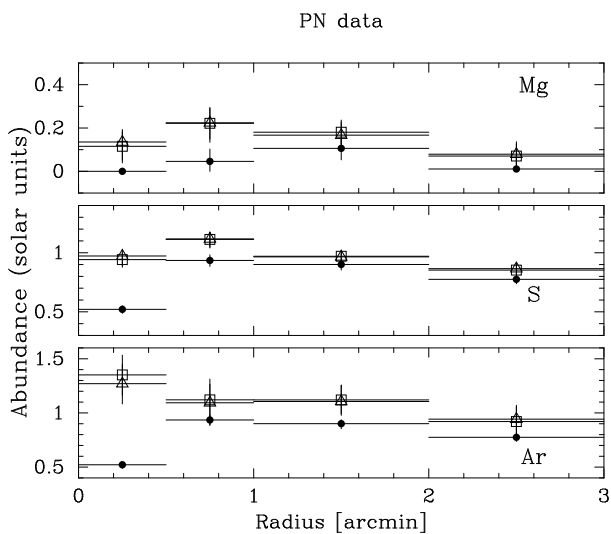}
      \caption{
PN profiles for Mg, S and Ar. Symbols as in Fig. 2. 
              }
   \end{figure}
%

Clearly single temperature models substantially 
underestimate the metal abundance in multi-temperature spectra.
This is not a new result, indeed something very similar  
has been observed in the ASCA spectra of galaxies and galaxy groups 
(e.g. Buote 2000a,b) and goes under the name 
of  ``$Fe$-bias''. 
The interested reader can find a detailed discussion of the 
``$Fe$-bia'' in Appendix A of Buote (2000a). 
The residuals of a single temperature model applied to
a multi temperature spectrum around the Fe-L line complex
(see fig. 5 of Buote 2000b) are also quite 
illustrative. 

Figure 1 tells us that models II and III behave very similarly 
in terms of their improvement with respect to model I.
In the outermost radial bin,  where we find the smallest
evidence for emission from more than a single temperature,
the abundances measured with models II and III
do not differ substantially from those derived from model I.
As we move inwards the evidence for  
multi-temperature gas becomes 
stronger, the $Fe$-bias becomes more important, and the abundance 
measurements with models II and III differ more and more from those
obtained with model I. 
In the innermost bin, the power-law emission from the nuclear sources 
contributes to raising the continuum with respect to the line 
emission thereby driving the abundances obtained with model I
further down. 

Since the metal abundances in the innermost region are sensitive
to the normalization of the power-law component
we have performed the following test to verify 
that this parameter is sufficiently well determined by the 
fitting procedure.
We have accumulated a spectrum for the innermost 
bin excising the emission from the nucleus
and knot A, using an exclusion radius of 7$^{\prime\prime}$ around 
the sources. We recall that for such a small exclusion radius 
the accumulated spectrum will still be contaminated by the
emission of the point-sources. We have then applied models II and
III to this spectrum. We find that the best fitting values for 
the metallicities are in agreement with those
derived by fitting the spectrum where the two sources
have not been eliminated. 

Close inspection of the $O$, $Si$, $Fe$ and $S$ profiles reveals a
small decrement in the innermost bin, the difference 
between the measurement in this region and the one in region 2 , which 
has the largest abundance, 
is significant at less than 2$\sigma$ for all elements in both MOS 
and PN with the exception of $Si$ for the PN, which is significant
at about 2.7$\sigma$. 

In summary, the metal abundance profiles derived by fitting single 
temperature models are very different from  the ones that are obtained 
by fitting the more appropriate spectral models II or III. The former 
all show a dramatic abundance decrease towards the center at radii 
smaller than one arcmin. The latter show a flattening of the 
abundance profile within 1 arcmin and a hint of a decrease
in the innermost bin. 
In the light of these results the evidence for absorption 
in the core of M87 is greatly reduced. 
Observations at higher angular resolution 
with the Chandra observatory will likely settle the issue 
definitively.

\begin{acknowledgements}
We thank the many people who have contributed to building, 
calibrating and operating the EPIC instrument on-board XMM-Newton.
S. Ghizzardi is thanked for useful discussions.
\end{acknowledgements}

\end{document}